\documentclass[conference]{IEEEtran}
\IEEEoverridecommandlockouts
\usepackage{cite}
\usepackage{amsmath,amssymb,amsfonts}
\usepackage{algorithmic}
\usepackage{graphicx}
\usepackage{textcomp}
\usepackage{multirow}
\usepackage{array}
\usepackage{tikz}
\makeatletter

\newcommand{\Rmnum}[1]{\expandafter\@slowromancap\romannumeral #1@}
\makeatother

\def\BibTeX{{\rm B\kern-.05em{\sc i\kern-.025em b}\kern-.08em
    T\kern-.1667em\lower.7ex\hbox{E}\kern-.125emX}}

\title{A Novel Framework for Visual Motion Imagery Classification Using 3D Virtual BCI Platform
\footnote{{\thanks{\hrule Research was partly supported by Institute of Information \& Communications Technology Planning \& Evaluation (IITP) grant funded by the Korea government (No. 2017-0-00432, Development of Non-Invasive Integrated BCI SW Platform to Control Home Appliances and External Devices by User’s Thought via AR/VR Interface) and partly funded by Institute of Information \& Communications Technology Planning \& Evaluation (IITP) grant funded by the Korea government (No. 2017-0-00451, Development of BCI based Brain and Cognitive Computing Technology for Recognizing User’s Intentions using Deep Learning).}
}
}}

\author{\IEEEauthorblockN{Byoung-Hee Kwon$^1$, Ji-Hoon Jeong$^1$, Dong-Joo Kim$^{1}$}
\IEEEauthorblockA{{$^1$Department of Brain and Cognitive Engineering, Korea University, Seoul, Republic of Korea} \\
bh\_kwon@korea.ac.kr, jh\_jeong@korea.ac.kr, dongjookim@korea.ac.kr}
}
\usepackage{enumerate}
\begin{document}

\maketitle

\begin{abstract}
In this study, 3D  brain-computer interface (BCI) training platforms were used to stimulate the subjects for visual motion imagery and visual perception. We measured the activation brain region and alpha-band power activity when the subjects perceived and imagined the stimuli. Based on this, 4-class were classified in visual stimuli session and visual motion imagery session respectively. The results showed that the occipital region is involved in visual perception and visual motion imagery, and alpha-band power is increased in visual motion imagery session and decreased in visual motion stimuli session. Compared with the performance of visual motion imagery and motor imagery, visual motion imagery has higher performance than motor imagery. The binary class was classified using one versus rest approach as well as analysis of brain activation to prove that visual-related brain wave signals are meaningful, and the results were significant.\\ 
\end{abstract}

\begin{small}\textbf{\textit{Keywords-brain-computer interface; visual motion imagery; electroencephalography; 3D BCI training plotform; robotic arm}\\}\end{small}

\section{Introduction}
Brain-computer interface (BCI) is a technology that enables direct communication between the user's brain and a computer and control of devices to reflect the user's intentions \cite{highertemporal}. Electroencephalography (EEG) equipment is widely used in BCI systems because it is non-invasive, and has high time resolution \cite{chen2016high}.

The BCI system controls the device according to the user's intention \cite{vaughan2003brain}, \cite{jeong2019trajectory} providing appropriate motor function for paralyzed patients. If the purpose of BCI is to control the device with user's intention \cite{wolpaw2002brain}, steady state visually evoked potential (SSVEP) \cite{won2015effect}, \cite{kwak2017convolutional}, evoked related potentials (ERPs) \cite{yeom2014efficient}, and motor imagery (MI) \cite{kim2014decoding}, \cite{kam2013non}, \cite{kim2016commanding} have used traditionally. However, these paradigms are difficult for users to intuitively accept. ERPs and SSVEP have an obvious limitation that brain signals should be generated through external stimuli and collected to identify user intentions. To compensate for these limitations, previous studies collected and analyzed brain signals that reflect user intentions using MI \cite{won2017motion}. MI is an endogenous-based paradigm and users imagine muscle movement. Accordingly, the area near sensory motor of brain that is related to movement is activated and this area is analyzed in various methods. In fact, many users imagined MI tasks in wrong way, and even if they are familiar with a proper way to imagine MI tasks, it is difficult to imagine muscle movement. MI also requires a long training time to be used in real online environments. This causes fatigue to users and hinders them in accepting high-quality brain signals. While the degree of freedom is very limited when controlling a device using MI, this study suggests the possibility of controlling a variety of high-level movements. 

Visual imagery allows users to control the device using internal imagination and allows for more intuitive imaginations than MI. Visual motion imagery can be classified based on analyzing the alpha wave areas. The intensity of alpha-band power varies when users are aware of visual stimuli and performing visual imaginations, which allows them to classify the two states \cite{sousa2017pure}. The potential of visual motion imagery as a control strategy in comparison to motor imagery was explored by Neuper et al\cite{neuper2005imagery}. 
In a recent study, Koizumi et al. clarified differences in brain activity between imagery and perception \cite{koizumi2018development}. Through this study, we have confirmed that perception and imagery are clearly distinct.

The purpose of this study is to compare the brain activation differences and activation regions between visual motion imagery and visual perception through EEG, and to classify each class in visual motion imagery. Many research groups have been investigated to control the arm prosthesis or the robotic arm based on BCI system. Meng et al. \cite{meng2016noninvasive} have investigated the feasibility of control a robotic arm in three-dimensional space using sensorimotor rhythm. However, the stimuli presented to users in previous studies were not three-dimensional stimuli and the imagery task was not intuitive.

\begin{figure*}[t!]
\centering
\centerline{\includegraphics[scale=1]{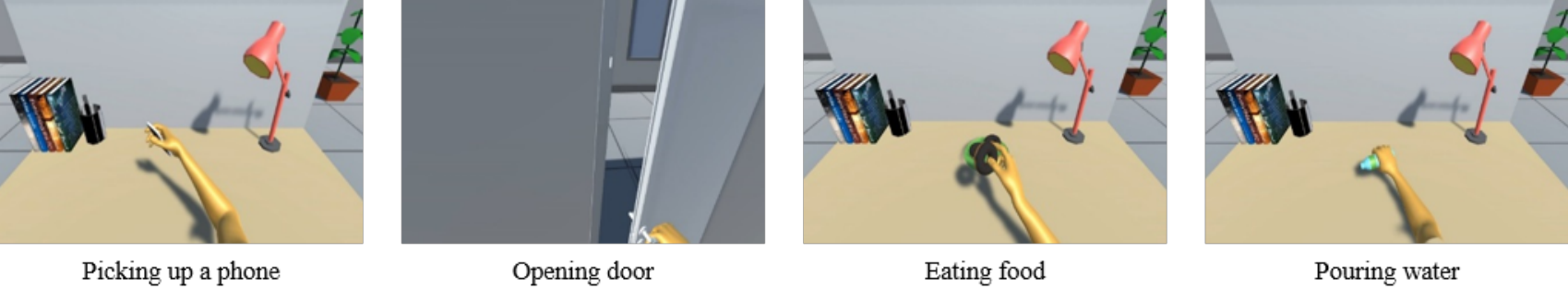}}
\caption{3D virtual BCI platform-based stimuli given to users in a perception session. Each stimulus is given to the user for five seconds and the movement is designed to have different meanings. The four above stimuli were shown only in the perception session, so that no aftereffect was left on the stimuli.}
\label{fig:res}
\end{figure*}   

\begin{figure}[t!]
\centering
\centerline{\includegraphics[width=\columnwidth]{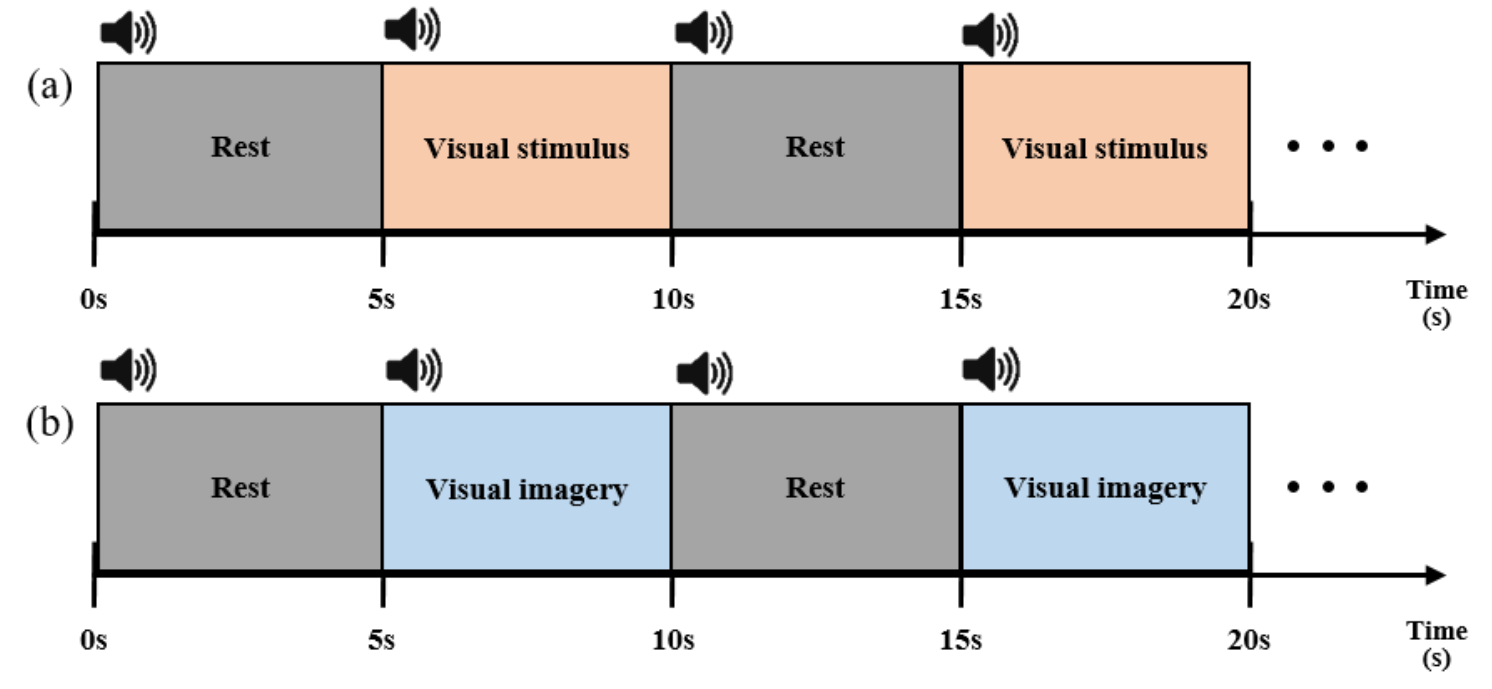}}
\caption{The paradigm of visual perception and visual motion imagery experiment. The user first conducted a session on the stimulus (a) and then a session on the imagery (b). An auditory stimulus of less than 0.5 seconds was given at the beginning of the rest phase and visual stimuli phase. Auditory stimuli were given corresponding auditory stimuli for each class.}
\label{fig:res}
\end{figure}   


Therefore, this study presented visual stimuli in three-dimensional space that users can intuitively accept and visual motion imagery as an intuitive imagination. We have hypothesized that provide more realistic stimuli to users results in better imaginations and can classify them. If the brain signals produced by visual imagery can be classified \cite{lee2015subject} as we have established a hypothesis, then users can control BCI-based robots such as robotic arms.\\

\section{Methods}

\subsection{Data Acquisition}
Five subjects between the ages of 25 and 30 participated in the experiment. The subjects who participated in the experiments were naïve BCI users. All participants were right-handed, and had normal vision and reported no medical or psychological disorders. We used BrainAmp amplifier and BrainVision Recorder for EEG data acquisition. EEG data were collected using 64 Ag/AgCl electrodes following 10-20 international system. The data were collected at 1,000 Hz only with 60 Hz notch filtering. The impedance of all electrodes were maintained below 10k$\Omega$ during the acquisitions. The protocols and environments were reviewed and approved by the Institutional Review Board at Korea University [1040548-KU-IRB-17-172-A-2].

\subsection{Experimental paradigm}
The visual stimulation paradigm was used as a guide to the visual motion imagery tasks. The video for visual stimulation was created using Unity3d and Blender (Blender 3D Engine: www.blender.org). Each video for stimulation consisted of eating food, pouring water, picking up a phone, and opening doors as illustrated in Fig. 1. The data of each session was controlled by MATLAB (MathWorks) using the Psychtoolbox (Psychtoolbox-3 distribution; http://www.psychtoolbox.org) \cite{kim2015improved}. The experiment consists of two sessions, a visual perception session, and a visual motion imagery session that take place on the same day (Fig. 2). In the visual stimuli session, the subject was stimulated for 5 seconds using each stimulation video. In an imagery session proceeded after a stimulation session, only a black screen is presented to the subject to minimize the side effects of visual stimulation. In visual motion imagery session, all subjects imagined each class according to the auditory cue of each class. In each session, the number of trials per class is 50. The subjects were instructed to minimize eye flickering and body movement during each session. The experiment was conducted in a dark environment so that the subjects could concentrate on perception and imagination, and the distance between the monitor and the subject's face was maintained at about 60 cm. 

\subsection{Data Analysis}
The data analysis was performed offline using BBCI toolbox and the EEGLAB toolbox (version 14.1.2b). To performance evaluation of visual motion stimuli and visual motion imagery, we used filtered EEG signals. Raw EEG signals filtered from 8 to 13 Hz with band-pass filter. The features of EEG signals should be extracted with common spatial pattern (CSP) \cite{channel}. Alpha band power was used in previous studies on visual motion imagery classification \cite{suk2011subject}. Thus, alpha band energy was extracted and filtered by CSP. To classify the visual motion stimuli and the visual motion imagery, we used the regularized linear discriminant (RLDA) as a classifier \cite{rus2017classification}.

EEG signals were band-pass filtered in [8-13] Hz, using a third order Butterworth filter for analyzing alpha-band power activity based on spatial information. Using BBCI toolbox, topographies of pre-processed data were visualized. We analyzed the spatial information of the visual stimuli data and visual motion imagery data in four time zones: (0$\sim$1,000) ms, (1,000$\sim$2,000) ms, (2,000$\sim$3,000) ms, and (3,000$\sim$4,000) ms.

In order to understand the variance of alpha-band power, we performed channel time-frequency of the imagery and stimulation data. Variation of spectral power according to the event-related changes at each time during the trials, and at each frequency, was analyzed using the event-related spectral perturbation (ERSP) method \cite{delorme2004eeglab}. ERSP analyses were performed for frequencies ranging from 3 to 50 Hz for all channels, using 200 time points. The baseline for calculating ERSP taken from the last 500 ms of the rest phase before imagery or stimuli phase.

\begin{figure*}[t!]
\centering
\centerline{\includegraphics[scale = 1.2, height=185pt]{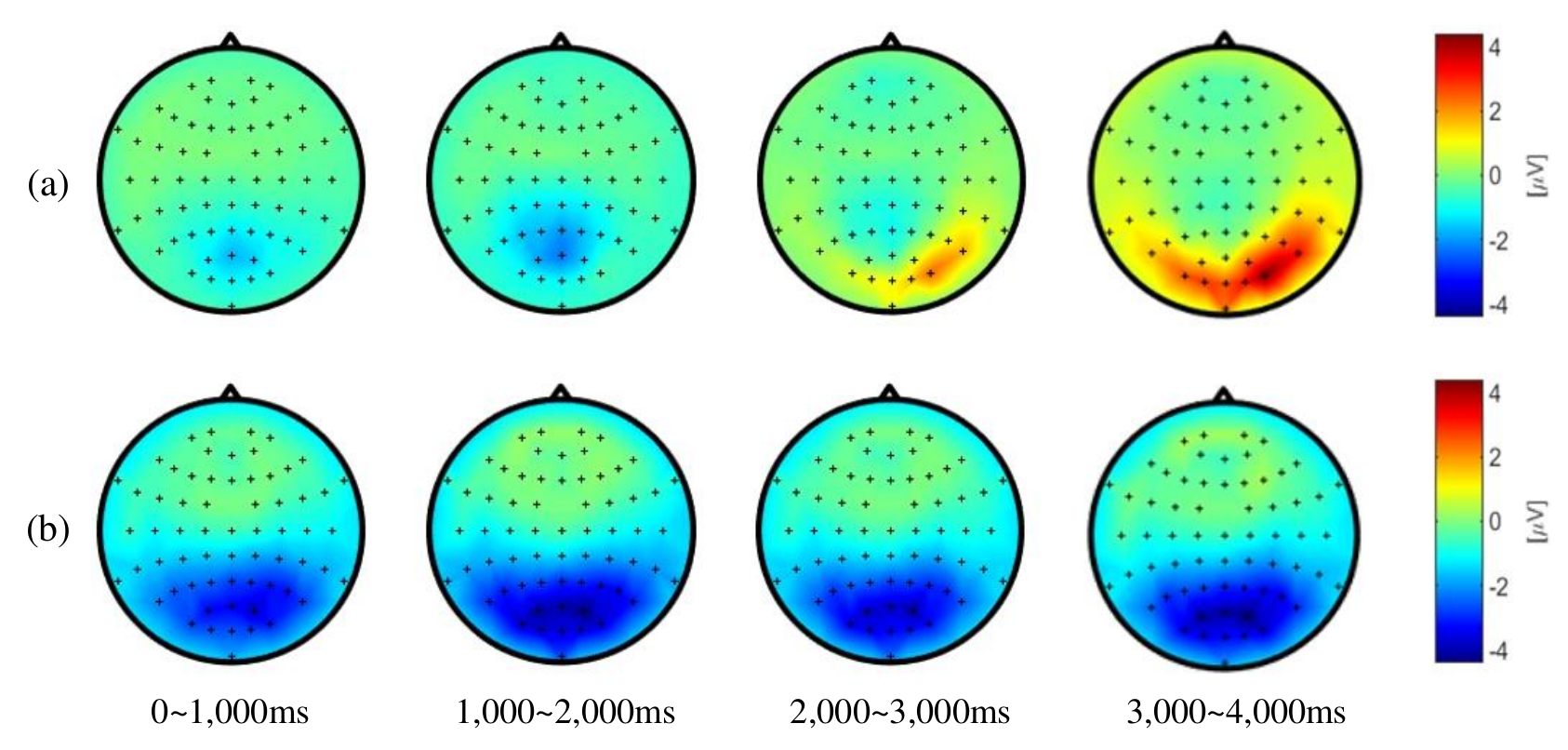}}
\caption{The activation of alpha wave in visual motion imagery and visual perception. (a) refers to the alpha wave activation in the visual imagery session and is activated near the occipital lobe over time. The following figure (b) shows the alpha wave activation in the visual perception session, and the power is decreased in the occipital lobe.}
\label{fig:res}
\end{figure*}

\begin{figure}[t!]
\centering
\centerline{\includegraphics[width=\columnwidth]{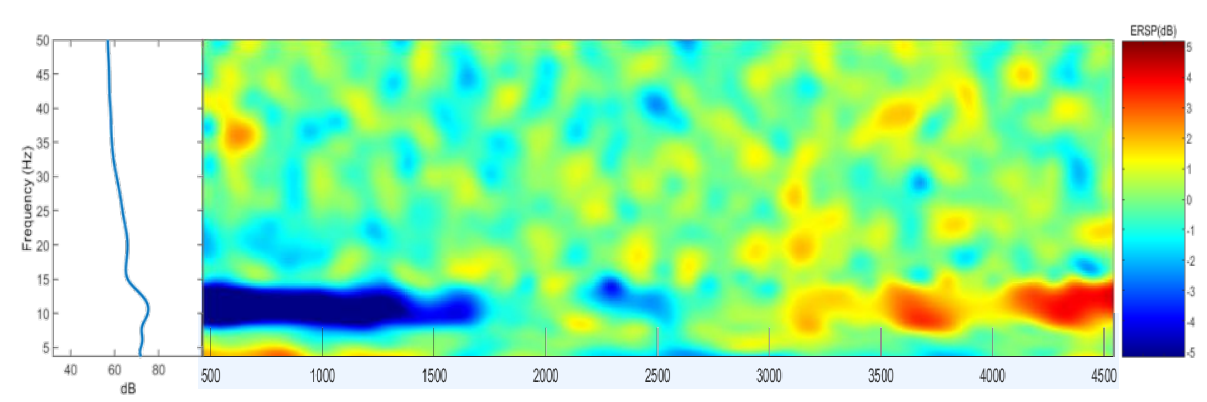}}
\caption{The event-related spatial perturbation of visual motion imagery. It shows that the power of alpha waves increases over time.}
\label{fig:res}
\end{figure}

To control BCI applications successfully, such as robotic arm control, we have attempted to classify \cite{zhu2016canonical} four imagery classes using a reduced number of channels and features. For analysis, we removed the effects of auditory stimuli before using the data. Since the auditory cue was given to the subjects during the visual motion imagery session, the initial 0.5 s was excluded after the auditory cue and a 0.5 to 4 s segment of the data was used for the analysis.\\

\section{Results and discussions}
\subsection{Neurophysiology Analysis}
Fig. 3 showed the visualization of the alpha-band power activity of each channel as spatial information. For the occipital channels' cluster, we found a significant main effect for the alpha-band power during visual motion imagery and visual perception session. The alpha-band power increase significantly during visual motion imagery session and decreased during visual perception session. When divided into 1,000ms intervals, the alpha-band power increased more clearly over time in the imagery session, and in the case of a perception session, the alpha-power band decreased regardless of time. This means that a characteristic brain signal was showed near the occipital region after a certain period of time after the onset of imagination and, in the case of perception, a characteristic brain signal was initiated directly from the occipital lobe because no preparatory step is required. Fig. 4 showed the ERSP of alpha-band power activity in visual motion imagery session using channel Oz. The graph on the left side of Fig. 4 showed the power distribution over the entire frequency range of the trial averaged data. The power increased in the alpha band, which means that characteristic brain signals evoked at frequencies related to visual imagery \cite{sousa2017pure},  \cite{koizumi2018development}. This result is consistent with the result that alpha band power increases in the occipital lobe over time when visual related imagery derived through Fig. 3. 

\subsection{Performance Evaluation}
Table \Rmnum{1} showed a comparison of the classification accuracies of visual related brain signals and motor related brain signals. The performance of visual motion imagery was 33.03\%, and the result was over 25\% chance level in four classes. The reason for the low visual motion imagery performance of sub4 is that user does not receive proper instruction, not a BCI illiteracy problem. This is because the performance for the visual perception of sub4 is much higher than that of other users, which means that it is possible to classify through visual-related brain waves. The results showed better performance than the motor imagery, which is commonly used in BCI-based control technologies such as robotic arms. We suggest the possibility of being used as an alternative application for BCI-based device control. Indeed, in a recent study, K. Koizumi and colleagues controlled drone through visual motion imagery in three-dimensional space and  yielded meaningful results \cite{kim2014decoding}. We suggest that through visual motion imagery, which is more intuitive than other paradigms, patients with disabilities can access BCI-based device control more easily.

Table \Rmnum{2} showed the performance comparisons to prove that subjects are imagining correctly in the visual imagery phase through one versus rest approach. Binary classification performances based on each class are 79.62\%, 81.36\%, 80.16\%, and 81.42\%, and it is an acceptable result. In one versus rest approach, there is little performance difference between classes, which means that the behavior or shape expressed in the class does not significantly affect classification. This can be inferred that an increase in the number of classes does not cause confusion among classes due to the confusion of visual motion. But as Table \Rmnum{1} shows, performance decreases as the number of classes increases. This is a critical problem when classifying multiple classes as machine learning in BCI domain. We expect to solve this problem using deep learning methods, and if the prediction is correct, we can increase the number of classes. This means that paralyzed patients can control BCI-based devices to suit their intentions.\\

\begin{table}[t!]
\small
\caption{The performance comparison of the executed session}
\renewcommand{\arraystretch}{1.3}
\resizebox{\columnwidth}{!}{%
\begin{tabular}{ccccccc}
\hline
\textbf{Task}                                                        & \textbf{Sub1}                                             & \textbf{Sub2}                                             & \textbf{Sub3}                                             & \textbf{Sub4}                                                           & \textbf{Sub5}                                             & \textbf{Average}                                          \\ \hline
\textbf{\begin{tabular}[c]{@{}c@{}}Visual\\ imagery\end{tabular}}    & \begin{tabular}[c]{@{}c@{}}42.80\%\\ (±1.36)\end{tabular} & \begin{tabular}[c]{@{}c@{}}33.15\%\\ (±1.76)\end{tabular} & \begin{tabular}[c]{@{}c@{}}31.85\%\\ (±0.85)\end{tabular} & \begin{tabular}[c]{@{}c@{}}26.40\%\\ (±0.99)\end{tabular}               & \begin{tabular}[c]{@{}c@{}}30.95\%\\ (±2.15)\end{tabular} & \begin{tabular}[c]{@{}c@{}}33.03\%\\ (±1.42)\end{tabular} \\
\textbf{\begin{tabular}[c]{@{}c@{}}Visual\\ perception\end{tabular}} & \begin{tabular}[c]{@{}c@{}}30.40\%\\ (±3.20)\end{tabular} & \begin{tabular}[c]{@{}c@{}}43.25\%\\ (±1.90)\end{tabular} & \begin{tabular}[c]{@{}c@{}}43.05\%\\ (±2.60)\end{tabular} & \begin{tabular}[c]{@{}c@{}}48.90\%\\ (±1.71)\end{tabular} & \begin{tabular}[c]{@{}c@{}}29.15\%\\ (±2.36)\end{tabular} & \begin{tabular}[c]{@{}c@{}}38.92\%\\ (±2.35)\end{tabular} \\ \hline
\textbf{\begin{tabular}[c]{@{}c@{}}Motor\\ imagery\end{tabular}}     & \begin{tabular}[c]{@{}c@{}}32.30\%\\ (±3.13)\end{tabular} & \begin{tabular}[c]{@{}c@{}}30.10\%\\ (±1.98)\end{tabular} & \begin{tabular}[c]{@{}c@{}}33.40\%\\ (±2.59)\end{tabular} & \begin{tabular}[c]{@{}c@{}}25.20\%\\ (±2.61)\end{tabular}               & \begin{tabular}[c]{@{}c@{}}33.35\%\\ (±2.31)\end{tabular} & \begin{tabular}[c]{@{}c@{}}30.87\%\\ (±7.18)\end{tabular} \\
\textbf{\begin{tabular}[c]{@{}c@{}}Motor\\ execution\end{tabular}}   & \begin{tabular}[c]{@{}c@{}}26.50\%\\ (±2.78)\end{tabular} & \begin{tabular}[c]{@{}c@{}}22.50\%\\ (±1.70)\end{tabular} & \begin{tabular}[c]{@{}c@{}}25.05\%\\ (±3.04)\end{tabular} & \begin{tabular}[c]{@{}c@{}}36.30\%\\ (±2.61)\end{tabular}               & \begin{tabular}[c]{@{}c@{}}27.25\%\\ (±2.79)\end{tabular} & \begin{tabular}[c]{@{}c@{}}25.52\%\\ (±2.58)\end{tabular} \\ \hline
\end{tabular}}
\end{table}
\begin{table}[t!]
\small
\caption{The result of one versus rest approach}
\renewcommand{\arraystretch}{1.3}
\resizebox{\columnwidth}{!}{%
\begin{tabular}{ccccccc}
\hline
\textbf{Task}                                                                  & \textbf{Sub1}                                           & \textbf{Sub2}                                           & \textbf{Sub3}                                           & \textbf{Sub4}                                           & \textbf{Sub5}                                           & \textbf{Average}                                        \\ \hline
\textit{\textbf{\begin{tabular}[c]{@{}c@{}}Eating\\ food\end{tabular}}}        & \begin{tabular}[c]{@{}c@{}}83.30\%\\ (±3.13)\end{tabular} & \begin{tabular}[c]{@{}c@{}}71.20\%\\ (±1.90)\end{tabular} & \begin{tabular}[c]{@{}c@{}}90.10\%\\ (±0.70)\end{tabular} & \begin{tabular}[c]{@{}c@{}}71.50\%\\ (±0.94)\end{tabular} & \begin{tabular}[c]{@{}c@{}}82.00\%\\ (±1.03)\end{tabular} & \begin{tabular}[c]{@{}c@{}}79.62\%\\ (±1.54)\end{tabular} \\ \hline
\textit{\textbf{\begin{tabular}[c]{@{}c@{}}Opening\\ door\end{tabular}}}       & \begin{tabular}[c]{@{}c@{}}82.30\%\\ (±2.01)\end{tabular} & \begin{tabular}[c]{@{}c@{}}79.30\%\\ (±3.37)\end{tabular} & \begin{tabular}[c]{@{}c@{}}82.70\%\\ (±0.90)\end{tabular} & \begin{tabular}[c]{@{}c@{}}73.90\%\\ (±2.16)\end{tabular} & \begin{tabular}[c]{@{}c@{}}88.60\%\\ (±1.16)\end{tabular} & \begin{tabular}[c]{@{}c@{}}81.36\%\\ (±1.92)\end{tabular} \\ \hline
\textit{\textbf{\begin{tabular}[c]{@{}c@{}}Picking up\\ a phone\end{tabular}}} & \begin{tabular}[c]{@{}c@{}}77.80\%\\ (±2.51)\end{tabular} & \begin{tabular}[c]{@{}c@{}}74.80\%\\ (±2.20)\end{tabular} & \begin{tabular}[c]{@{}c@{}}87.20\%\\ (±0.94)\end{tabular} & \begin{tabular}[c]{@{}c@{}}71.60\%\\ (±1.49)\end{tabular} & \begin{tabular}[c]{@{}c@{}}89.40\%\\ (±1.58)\end{tabular} & \begin{tabular}[c]{@{}c@{}}80.16\%\\ (±1.74)\end{tabular} \\ \hline
\textit{\textbf{\begin{tabular}[c]{@{}c@{}}Pouring\\ water\end{tabular}}}      & \begin{tabular}[c]{@{}c@{}}80.90\%\\ (±2.27)\end{tabular} & \begin{tabular}[c]{@{}c@{}}79.20\%\\ (±1.16)\end{tabular} & \begin{tabular}[c]{@{}c@{}}84.70\%\\ (±0.70)\end{tabular} & \begin{tabular}[c]{@{}c@{}}76.40\%\\ (±0.70)\end{tabular} & \begin{tabular}[c]{@{}c@{}}85.90\%\\ (±1.45)\end{tabular} & \begin{tabular}[c]{@{}c@{}}81.42\%\\ (±1.12)\end{tabular} \\ \hline
\end{tabular}}
\end{table}

\section{Conclusion and Future Work}
In this study, brain signals related to visual perception and visual motion imagery were analyzed by brain region and frequency. And we classified visual related imagery using 3D BCI training platform. As a result, performance exceeded the chance level for all classes. In addition, visual motion imagery brings out a better performance than motor imagery and proves that a solid visual imagination is possible through a one versus rest approach. However, for application, the performance of visual motion imagery (33.03\% based on four classes in this study) is still low. We can present the possibility that there is a deep learning architecture suitable for classification rather than machine learning for EEG classification, and it can improve future performance.

In Fig. 4, we can see a similar aspect to event-related desynchronization (ERD) and event-related synchronization (ERS) in motor imagery. We are going to analyze neurophysiological analysis of what this means in visual imagery.

After questioning the inconvenience to the subjects about the experiment, the paradigm proposed in this study had the problem that each class's memory becomes blurred over time in the visual motion imagery session. Each session has 50 trials per class, and subjects should perform a total of 200 trials in one session. Thus, in the visual motion imagery session where no visual stimulus is given, the memory of the stimuli seen in the previous visual perception session is dimmed. Also, the paradigm divided into both different sessions can give fatigue and pressure to subjects, so we should develop a more concise and effective paradigm consisting of sessions made by stimuli and imagery. A comfortable environment can be created, which can improve the performance by acquiring good brain signals.\\

\section{ACKNOWLEDGMENT}
The authors thanks to J.-H. Cho, B.-H. Lee, and D.-Y. Lee for setting experiment environment and K.-H Shim for developing the experimental codes.\\

\bibliographystyle{IEEEbib}
\bibliography{refs}


\end{document}